\documentclass[runningheads]{llncs}
\usepackage[T1]{fontenc}

\usepackage{xcolor}
\usepackage{graphicx}
\usepackage{relsize}

\usepackage{color}
\usepackage{colortbl}

\newcommand{\dgc}{0.8}
\newcommand{\lgc}{0.9}
\definecolor{dg}{rgb}{\dgc, \dgc, \dgc}
\definecolor{lg}{rgb}{\lgc, \lgc, \lgc}

\newcommand{\Wic}{\cellcolor[gray]{\lgc}}
\newcommand{\capBox}[2]{%
  \begingroup\setlength{\fboxsep}{0.8pt}%
  \colorbox{#2}{\texttt{\hspace*{0.5pt}\vphantom{Ay}\footnotesize #1\hspace*{0.5pt}}}%
  \endgroup
}

\usepackage{float}
\usepackage{multirow} 

\usepackage{hyperref}
\hypersetup{
colorlinks   = true,
citecolor    = blue,
urlcolor    =  blue
}
\usepackage{amsmath} 

\usepackage{booktabs}
\usepackage{makecell}
\usepackage{ragged2e}
\usepackage{tabularx}
\newcolumntype{L}{X}
\newcolumntype{C}{>{\centering\arraybackslash}X}
\newcolumntype{R}{>{\raggedleft\arraybackslash}X}
\newcommand{\footURL}[1]{\footnote{\url{#1}}}

\usepackage[numbers,square,sort&compress]{natbib}
\usepackage{todonotes}

\newcommand{\ex}[1]{\Wic #1\phantom{~[00]}}

\begin{document}
\title{From Monolith to Microservices: A Comparative Evaluation of Decomposition Frameworks}

\author{
  Mineth Weerasinghe\inst{1}\orcidID{0009-0003-6427-6794} \and
  Himindu Kularathne\inst{1}\orcidID{0009-0008-3794-3260} \and
  Methmini Madhushika\inst{1}\orcidID{0009-0004-9566-8536} \and
  Danuka Lakshan\inst{1}\orcidID{0009-0002-8864-9957} \and 
  Nisansa de Silva\inst{1}\orcidID{0000-0002-5361-4810} \and
  Adeesha Wijayasiri\inst{1}\orcidID{0000-0002-1807-3545} \and
  Srinath Perera\inst{2}\orcidID{0000-0002-4457-903X}
}

\authorrunning{M. Weerasinghe et al.}

\institute{
Department of Computer Science \& Engineering,
University of Moratuwa,\\Sri Lanka\\
\email{\{mineth.21,himindu.21,methmini.21,dhanuka.21,NisansaDdS,adeeshaw\}@cse.mrt.ac.lk}
 \and
WSO2 LLC\\ 
\email{srinath@wso2.com}\\
}

\maketitle              
\begin{abstract}
Software modernisation through the migration from monolithic architectures to microservices has become increasingly critical, yet identifying effective service boundaries remains a complex and unresolved challenge. 
Although numerous automated microservice decomposition frameworks have been proposed, their evaluation is often fragmented due to inconsistent benchmark systems, incompatible metrics, and limited reproducibility, thus hindering objective comparison.
This work presents a unified comparative evaluation of state-of-the-art microservice decomposition approaches spanning static, dynamic, and hybrid techniques. Using a consistent metric computation pipeline, we assess the decomposition quality across widely used benchmark systems (\texttt{JPetStore}, \texttt{AcmeAir}, \texttt{DayTrader}, and \texttt{Plants}) using Structural Modularity (SM), Interface Number (IFN), Inter-partition Communication (ICP), Non-Extreme Distribution (NED), and related indicators. Our analysis combines results reported in prior studies with experimentally reproduced outputs from available replication packages.
Findings indicate that the hierarchical clustering-based methods, particularly HDBScan, produce the most consistently balanced decompositions across benchmarks, achieving strong modularity while minimizing communication and interface overhead.

\end{abstract}

\section{Introduction}
The adoption of microservice architecture has revolutionized modern software engineering by enabling scalability, flexibility, and independent deployment~\cite{dragoni2017microservices,newman2021building}. Recent comparative studies confirm that while monolithic systems often suffer from performance bottlenecks when scale increases, microservice architectures deployed on containers can achieve lower response times, improved throughput, and better resource utilization, despite introducing challenges such as orchestration and higher infrastructure complexity~\cite{tapia2020monolithic}. However, transforming legacy monolithic systems into well-structured microservices remains a challenging task. In this context, decomposition refers to the process of breaking a monolithic application into smaller, autonomous microservices with clear functional boundaries \cite{mazlami2017microservice}.  Among the most critical steps in this transformation is identifying appropriate service boundaries, a process that directly impacts modularity, maintainability, and system performance~\cite{al2021semantic}.

Recent literature reveals that most automated decomposition systems follow a comparable workflow comprising input collection, monolith analysis, service identification, and representation or RESTification stages~\cite{mazlami2017microservice}. For instance, CARGO~\cite{nitin2022cargo} employs static program dependency analysis to construct a context-sensitive system graph and applies label propagation to refine service partitions. Similarly, MONO2REST~\cite{lecrivain2025mono2rest} automates service identification and exposure through a reusable RESTification pipeline. Graph-clustering-based approaches such as the combinatorial optimization method~\cite{filippone2023graphcluster} also rely on code-structure modeling and dependency analysis to guide boundary extraction. Together, these studies illustrate a systematic pipeline that transforms legacy monoliths into microservice-oriented architectures through data-driven analysis and modular service extraction.

Despite the rapid progress in automated microservice decomposition, a significant gap remains in how existing approaches are evaluated and compared. Prior studies often introduce new techniques but rely on inconsistent benchmark systems (e.g. \cite{fosci2020, mathai2021chgnn}), incompatible metric definitions (e.g. \cite{fosci2020, sellami2022hierarchical}), or selective reporting of results (e.g. \cite{mazlami2017microservice, mathai2021chgnn}), making it difficult to understand the true strengths and limitations of each method. Moreover, many tools depend on either purely structural (static analysis)~\cite{mitchell2008} or purely execution (dynamic analysis)~\cite{fosci2020} signals, overlooking the fact that real-world service boundaries emerge from a combination of code dependencies, domain semantics, and architectural intent~\cite{saied2024migration}. Therefore, developers have no clear idea about which tool should be used for different monolithic structures, levels of system complexity, or domain-driven design expectations ~\cite{taibi2021empirical,saied2024migration}. This lack of unified evaluation, reproducible experimentation, and cross-framework comparison motivated us to undertake a systematic, tool-agnostic assessment of modern decomposition frameworks.

In this work, we conduct a unified and reproducible evaluation of several state-of-the-art microservice decomposition tools across widely used benchmark systems. Using a consistent metric computation pipeline, we compare static, dynamic and hybrid approaches, including Bunch, MEM~\cite{mazlami2017microservice}, FoSCI~\cite{fosci2020}, CoGCN~\cite{desai2021gnn}, Mono2Micro~\cite{kalia2021mono2micro}, HDBScan~\cite{sellami2022hierarchical}, a-BMSC~\cite{saied2024migration}, CHGNN~\cite{mathai2021chgnn} and MonoEmbed~\cite{sellami2025contrastive}, over multiple quality indicators such as Structural Modularity (SM)~\cite{fosci2020}, Interface Number (IFN)~\cite{fosci2020}, Inter-partition Communication (ICP)~\cite{kalia2020mono2micro}, and Non-Extreme Distribution (NED)~\cite{saied2024migration}.

The key contributions of this study are as follows:
\begin{itemize}
\renewcommand\labelitemi{$\bullet$}
    \item We curate a set of benchmark systems and evaluation metrics commonly used in microservice decomposition frameworks, and present results from existing studies. 
    \item To address gaps in prior work, we run experiments for dataset–model pairings that have not been previously studied.
    \item We conduct a systematic comparative analysis across all collected and produced results, enabling the identification of the most effective tools for microservice extraction.
\end{itemize}

\section{Background and Related Work}

We begin by describing the microservice identification tools, followed by the evaluation metrics used to compare them, and finally discuss the common benchmark datasets applied in prior studies.

\subsection{Microservice Identification Approaches}

Microservice identification techniques can be broadly classified into artefact-driven, static, dynamic and hybrid approaches~\cite{mohottige2025reengineering}.

\textbf{Artifact-driven} approach uses requirements, design diagrams, UML diagrams, data-flow diagrams, business processes, use cases, user stories, domain models, and other design artifacts to identify bounded contexts. These bounded contexts are then treated as microservices. RapidMS~\cite{zhang2023rapidms} is an example for this approach; however, since it is requirement driven and not compatible with our code based evaluation metrics, we do not consider this approach in our comparative analysis. 

\textbf{Static analysis} analyzes source-code dependencies (e.g., inheritance), database schema relations, and the history of source-code repositories such as commits. This information is used to identify potential service boundaries. This is the most common and straightforward approach in microservice identification. Bunch~\cite{mitchell2008}, MEM~\cite{mazlami2017microservice}, CoGCN~\cite{desai2021gnn}, HDBScan~\cite{sellami2022hierarchical}, a-BMSC~\cite{saied2024migration}, and MonoEmbed~\cite{sellami2025contrastive} discussed in our work are examples of this approach.

\textbf{Dynamic analysis} uses runtime information such as monitoring data, execution time correlations, and system-generated logs to identify microservice boundaries. Since it requires runtime behaviour, the monolithic system must be executed or simulated to collect this information. We discuss FoSCI~\cite{fosci2020} and Mono2Micro~\cite{kalia2021mono2micro} as systems that follow this approach.

\textbf{Hybrid approaches} combine the static and dynamic methods discussed above. Typically, a primary approach is used together with features from another to reduce the limitations of the first. For example, a static-analysis approach can incorporate execution traces to make more accurate decisions about service boundaries. CHGNN~\cite{mathai2021chgnn} covered in our study is a hybrid microservice decomposition method where static analysis, such as call relationships, CRUD interactions, and inheritance, are integrated with dynamic behavioural analysis derived from execution traces. 

\subsection{Evaluation Metrics for Service Decomposition}
\label{sec:eval}

The quality of a microservice decomposition is often evaluated using quantitative metrics that assess structural modularity, functional cohesion, and service distribution balance. The most commonly adopted metrics across existing studies include Structural Modularity (SM), Inter-partition Communication (ICP), Business Capability Purity (BCP), Interface Number (IFN), and Non-Extreme Distribution (NED).

\subsubsection{Structural Modularity (SM)~\cite{fosci2020}}
measures the modularity quality of partitions as the structural cohesiveness of classes within a partition and the coupling between partitions. It is computed as shown in Equation~\ref{eq:sm}, where M
denotes the total number of partitions. In the equation, $scoh_i = \frac{\mu{i}}{m_i^2}$ represents the cohesiveness within partition $i$, with ${\mu{i}}$ denoting the number of calls within
the partition and ${m_i}$ denoting the number of classes in partition $i$. Similarly, $scop_{i,j} = \frac{\gamma_{i,j}}{2(m_i*m_j)}$ represents the coupling between partitions $i$ and $j$, where
${\gamma{i,j}}$ denotes the number of calls made between them. Higher SM values indicate better modular decomposition~\cite{mitchell2008}.
\setlength{\abovedisplayskip}{3pt}
\setlength{\belowdisplayskip}{3pt}
\begin{equation}
    SM = \frac{1}{M}\sum_{i=0}^{M}scoh_i - \frac{1}{(M(M-1))/2}\sum_{i\neq j}^{M}scop_{i,j}
    \label{eq:sm}
\end{equation}

SM is widely recognized as the \textit{primary indicator of decomposition quality}, capturing the extent to which cohesion is preserved within services while minimizing inter-service coupling~\cite{mitchell2008, filippone2023graphcluster, saied2024migration}.
 Higher SM values indicate that cohesion is preserved and inter-service coupling is minimized, supporting the single responsibility principle, which is fundamental in microservice design. Therefore, SM often serves as the main optimization objective in decomposition frameworks.

\subsubsection{Interface Number (IFN)~\cite{fosci2020}} measures the average number of interfaces per microservice and is calculated as shown in Equation~\ref{eq:ifn}, where $ifn_i$ is the number of interfaces in the $i^{th}$ microservice. Lower IFN values indicate simpler, less fragmented services~\cite{saied2024migration}.

\setlength{\abovedisplayskip}{3pt}
\setlength{\belowdisplayskip}{3pt}
\begin{equation}
IFN = \frac{1}{N}\sum_{i=1}^{N}ifn_i
\label{eq:ifn}
\end{equation}

IFN provides insight into service complexity. A higher number of exposed interfaces suggests fragmentation and potential redundancy. Thus, minimizing IFN helps maintain clarity in API boundaries and simplifies integration across services.

\subsubsection{Inter-partition Communication (ICP)~\cite{kalia2020mono2micro}}
quantifies the percentage of runtime calls occurring between two partitions, computed as shown in Equation~\ref{eq:icp}, where ${c_{i,j}}$ denotes the number of calls between partitions $i$ and $j$.  Lower ICP values indicate better separation among services.

\setlength{\abovedisplayskip}{3pt}
\setlength{\belowdisplayskip}{3pt}
\begin{equation}
ICP_{i,j} = \frac{c_{i,j}}{\sum_{i=1}^{M}\sum_{j=1, j\neq i}^{M}c_{i,j}}
\label{eq:icp}
\end{equation}

ICP directly measures the amount of cross-service interaction at runtime. Excessive communication between services increases latency and reduces autonomy, undermining the benefits of microservices. Hence, lower ICP values are critical to achieving independently deployable and loosely coupled services~\cite{bolanowski2022efficiency}

\subsubsection{Non-Extreme Distribution (NED)~\cite{saied2024migration}} assesses how evenly service sizes are distributed within a decomposition. It is defined as shown in Equation~\ref{eq:ned}, where lower NED values reflect more balanced microservice size distributions. Originally proposed by~\citet{wu2005comparison}, this metric identifies disproportionate service partitions where a few microservices dominate in size. Here N refers to the total number of microservices, and $n_k$ refers to the number of non-extreme
microservices. A microservice is considered non-extreme when the number of classes it contains satisfies $5 \le |k| \le 20$~\cite{kalia2021mono2micro,scanniello2010architectural}, where $k$ denotes the size of microservice.  
\setlength{\abovedisplayskip}{3pt}
\setlength{\belowdisplayskip}{3pt}
\begin{equation}
NED = 1- \frac{\sum_{k=0}^{N} n_k}{|N|}
\label{eq:ned}
\end{equation}

NED evaluates size balance among microservices. Extremely uneven service distributions, where a few services dominate, can lead to scalability bottlenecks. Hence, maintaining moderate NED values ensures that services remain evenly balanced and maintain granularity consistent with the ``small and independent'' nature of microservices~\cite{dragoni2017microservices}.

Among the evaluated metrics, ICP and IFN also significantly affect communication efficiency and system performance.
SM and ICP together capture the essential balance between cohesion and coupling, while IFN and NED complement them by revealing interface simplicity and service size balance. The relationship between modularity and runtime efficiency has also been empirically validated by~\citet{tapia2020monolithic}, who compared equivalent monolithic and microservice implementations under stress-testing scenarios. Their regression-based analysis demonstrated that modular decomposition (analogous to higher SM) significantly improves resource utilization and scalability, while excessive service fragmentation increases orchestration overhead and inter-service latency effects captured quantitatively by ICP.
\subsection{Benchmark Datasets and Experimental Frameworks}

For evaluating microservice decomposition techniques, a consistent set of benchmark systems is essential to ensure reproducibility and fair comparison across tools. Most studies employ open-source, medium-scale monolithic applications~\cite{abgaz2023decomposition} that capture realistic enterprise structures while remaining manageable for analysis.
In this work, we have used 4 benchmark systems (Table~\ref{tab:combined_results})
\texttt{JPetStore}\footURL{https://github.com/mybatis/jpetstore-6},
\texttt{AcmeAir}\footURL{https://github.com/acmeair/acmeair},
\texttt{DayTrader}\footURL{https://github.com/WASdev/sample.daytrader7} and
\texttt{Plants}. 

\texttt{JPetStore} is a Java-based e-commerce application commonly used to evaluate clustering-based decomposition techniques due to its layered architecture and moderate codebase size. \texttt{AcmeAir}, developed by IBM, is a cloud-oriented airline reservation system that models realistic business transactions whose design makes it suitable for assessing decomposition tools that rely on dynamic traces or runtime interaction data. \texttt{DayTrader}~\citep{daytrader2007} is a stock trading application developed by IBM, which provides realistic transaction-heavy workloads, making it useful for evaluating decomposition methods under high-throughput enterprise scenarios. 
\texttt{Plants} is a Java-based application modelling an online plant nursery system with modules for catalogue browsing, cart management, and order processing. The version commonly used in research is a simplified variant derived from PlantsByWebSphere\footURL{https://github.com/WASdev/sample.plantsbywebsphere}, with Java EE components removed for easier analysis. Due to methodological incompatibilities with our evaluation pipeline, which targets monolith-to-microservice decomposition and relies on a uniform static and/or hybrid metric computation framework (SM, IFN, ICP, NED), benchmark datasets such as \texttt{TrainTicket}\footURL{https://github.com/FudanSELab/train-ticket}, \texttt{Spring PetClinic}\footURL{https://github.com/spring-projects/spring-petclinic}, and \texttt{SockShop}\footURL{https://github.com/microservices-demo/microservices-demo} are not included in this study.
\texttt{TrainTicket}~\citep{trainticket} is a microservice-based train booking system with publicly available ground-truth service boundaries, enabling objective comparison of decomposition techniques.
\texttt{Spring PetClinic} is a lightweight, domain-driven sample application featuring veterinary clinic management workflows, commonly used to assess tools on smaller, well-structured codebases.
\texttt{SockShop}, is a reference microservice application e-commerce application, frequently used to validate service identification and communication patterns due to its representative microservice architecture.

\section{Evaluation Methodology}

All experiments were executed on multiple developer workstations rather than a single controlled server environment. Since our evaluation involved running each decomposition tool independently on the above benchmark systems, the computations were distributed across workstations with comparable but not identical hardware specifications.As a result, runtime performance, execution latency, and throughput are considered out of scope in this study. This setup reflects a realistic practitioner-oriented scenario in which decomposition tools are typically executed on commodity hardware rather than specialised clusters. Because the goal of this work is to evaluate decomposition \emph{quality} rather than raw execution time, none of the tools used in this study rely on hardware-accelerated operations or GPU-based learning phases. 
We evaluated the results of the decomposition tools by comparing them against the evaluation metrics we mentioned in section~\ref{sec:eval}.

We locally executed the tools CHGNN and MonoEmbed, using the open-source benchmarks (JPetStore, AcmeAir, DayTrader, and Plants). 
For CHGNN, we relied on the source code and replication package given by authors, and used the default configuration, which gives 30 decompositions. CHGNN results are not reported for \texttt{JPetStore} due to missing input artifacts in the replication package.
We relied on the replication package provided by the authors to execute the decomposition process as originally defined for MonoEmbed as well.
Each decomposition output (in either JSON or database form) was post-processed through a unified metric computation pipeline to obtain the evaluation values. For the remaining tools, the metric values were taken directly from the corresponding research papers, as no executable artifacts or replication packages were available.

Finally, to summarise the overall decomposition quality across the individual metrics, we compute an aggregate score for each tool and benchmark.  For each metric $m \in \{\mathrm{SM}, \mathrm{IFN}, \mathrm{ICP}, \mathrm{NED}\}$ and tool $t$, we standardise the raw values across all tools on the same benchmark using $z$-score normalisation,
where $x_{m,t}$ is the raw value for metric $m$, and $\mu_{m}$ and $\sigma_{m}$ are, respectively, the mean and standard deviation of $m$ over all tools for that benchmark. We then follow the weighting rationale proposed by~\citet{sellami2025contrastive} in their evaluation of MonoEmbed to aggregate the component results into a composite score as shown in Equation~\ref{eq:overall-score}, where $w_m$ indicates the weight given for metric $m$.

\setlength{\belowdisplayskip}{3pt}
\begin{equation}
  \label{eq:overall-score}
  \mathrm{Score}(T)
  = \frac{ \mathlarger{\sum} w_m \big( \frac{x_{m,t} - \mu_{m}}{\sigma_{m}} \big)
    }{\mathlarger{\sum} |w_m|}.
\end{equation}

When selecting $W$, we follow the convention of~\citet{sellami2025contrastive} such that $\Sigma w_m = 0$, where the metrics desired to be increased are given positive weights that are then balanced by the negative weights given to the metrics that are desired to be decreased. Further, they have indicated that among the negative weights, NED should be considered half as valuable as measures of coupling between the microservices. However, unlike~\citet{sellami2025contrastive}, we evaluate both IFN and ICP to measure the said coupling with definitions related to one another. Thus, the three negative weights end up being $-1$ each. Finally, as mentioned before, to balance the negative metrics, the weight for SM is set at $3$. Thus, the weight vector is $W=\{3,-1,-1,-1\}$.

\section{Results and Discussion}

\begin{table}[!] 
\centering
\caption{Comparison of decomposition tools across different projects: Structural Modularity (SM), Inter-partition Communication (ICP), Interface Number (IFN), Non-Extreme Distribution (NED), and the number of microservices (Micro). All values for \textit{a-BMSC} are taken from \cite{saied2024migration}, while results for baseline methods are from \cite{kalia2021mono2micro, mathai2021chgnn, mazlami2017microservice, fosci2020, mitchell2008, hierdecomp2019}. Values in the cells shaded in \capBox{\strut light gray}{lg} are from experiments or calculations carried out by us.} \label{tab:combined_results}
\resizebox{!}{0.38\textheight}{
\begin{tabularx}{\textwidth}{|l|l|R|R|R|R|R|R|}
\hline \textbf{Data Set} & \textbf{Tool} & \textbf{SM}$\uparrow$ & \textbf{IFN}$\downarrow$ & \textbf{ICP}$\downarrow$ & \textbf{NED}$\downarrow$ & \textbf{Micro} & \textbf{Score}$\uparrow$ \\ \hline 
\multirow{10}{*}{DayTrader}  
& Bunch~\cite{mitchell2008} & 0.18~\cite{saied2024migration} & 11.00~\cite{saied2024migration} & 0.50~\cite{saied2024migration} & 0.65~\cite{saied2024migration} & 5~\cite{saied2024migration} & \Wic  -0.58\\ 
& MEM~\cite{mazlami2017microservice} & 0.32~\cite{saied2024migration} & 3.50~\cite{saied2024migration} & 0.25~\cite{saied2024migration} & 0.95~\cite{saied2024migration} & 18~\cite{saied2024migration} &  \Wic 0.04 \\ 
& FoSCI~\cite{fosci2020} & 0.30~\cite{saied2024migration} & 7.00~\cite{saied2024migration} & 0.82~\cite{saied2024migration} & 0.55~\cite{saied2024migration} & 19~\cite{saied2024migration} & \Wic -0.17 \\ 
& CoGCN~\cite{desai2021gnn} & \textbf{0.49}~\cite{saied2024migration} & 3.00~\cite{saied2024migration} & 0.35~\cite{saied2024migration} & 0.70~\cite{saied2024migration} & 19~\cite{saied2024migration} & \Wic \textbf{0.73} \\ 
& Mono2Micro~\cite{kalia2021mono2micro} & 0.08~\cite{kalia2021mono2micro} & 1.92~\cite{kalia2021mono2micro} & 0.35~\cite{kalia2021mono2micro} & \textbf{0.34}~\cite{kalia2021mono2micro} & 19~\cite{saied2024migration} & \Wic -0.15 \\ 
& HDBScan~\cite{sellami2022hierarchical} & 0.30~\cite{sellami2022hierarchical} & \textbf{0.05}~\cite{sellami2022hierarchical} & \textbf{0.05}~\cite{sellami2022hierarchical} & 0.70~\cite{sellami2022hierarchical} & \ex{-} & \Wic 0.47 \\
& a-BMSC~\cite{saied2024migration} & 0.43~\cite{saied2024migration} & 1.20~\cite{saied2024migration} & 0.63~\cite{saied2024migration} & 0.65~\cite{saied2024migration} & 23~\cite{saied2024migration} & \Wic 0.49 \\ 
& CHGNN~\cite{mathai2021chgnn} & \ex{0.13} & \ex{5.70} & \ex{0.55} & \ex{0.50} & \ex{6} & \Wic -0.42 \\ 
& MonoEmbed~\cite{sellami2025contrastive} & \ex{0.13} & \ex{1.35} & \ex{0.59} & \ex{0.66} & \ex{29} & \Wic -0.40 \\
\hline 

\multirow{10}{*}{Plants} 
& Bunch~\cite{mitchell2008} & 0.22~\cite{saied2024migration} & 5.50~\cite{saied2024migration} & 0.30~\cite{saied2024migration} & 0.10~\cite{saied2024migration} & 3~\cite{saied2024migration} & \Wic -0.18 \\ 
& MEM~\cite{mazlami2017microservice} & 0.54~\cite{saied2024migration} & 3.30~\cite{saied2024migration} & 0.22~\cite{saied2024migration} & 0.25~\cite{saied2024migration} & 6~\cite{saied2024migration} & \Wic 0.71 \\ 
& FoSCI~\cite{fosci2020} & 0.33~\cite{saied2024migration} & 4.00~\cite{saied2024migration} & 0.47~\cite{saied2024migration} & 0.70~\cite{saied2024migration} & 7~\cite{saied2024migration} & \Wic -0.20 \\ 
& CoGCN~\cite{desai2021gnn} & 0.33~\cite{saied2024migration} & 3.50~\cite{saied2024migration} & 0.55~\cite{saied2024migration} & 0.50~\cite{saied2024migration} & 8~\cite{saied2024migration} & \Wic -0.13 \\ 
& Mono2Micro~\cite{kalia2021mono2micro} & 0.08~\cite{kalia2021mono2micro} & 6.00~\cite{kalia2021mono2micro} & 0.38~\cite{kalia2021mono2micro} & \textbf{0.04}~\cite{kalia2021mono2micro} & 7~\cite{saied2024migration} & \Wic -0.57 \\ 
& HDBScan~\cite{sellami2022hierarchical}& \textbf{0.60}~\cite{sellami2022hierarchical} & \textbf{1.00}~\cite{sellami2022hierarchical} & \textbf{0.03}~\cite{sellami2022hierarchical} & 0.80~\cite{sellami2022hierarchical} &\ex{-} & \Wic \textbf{0.93} \\ 
& a-BMSC~\cite{saied2024migration} & 0.49~\cite{saied2024migration} & 1.80~\cite{saied2024migration} & 0.68~\cite{saied2024migration} & 0.75~\cite{saied2024migration} & 14~\cite{saied2024migration} &  \Wic 0.17 \\ 
& CHGNN~\cite{mathai2021chgnn} & \ex{0.17} & \ex{3.60} & \ex{0.51} & \ex{0.20} & \ex{5} & \Wic -0.33 \\ 
& MonoEmbed~\cite{sellami2025contrastive} & \ex{0.11} & \ex{1.55} & \ex{0.26} & \ex{0.82} & \ex{11} & \Wic -0.39 \\ \hline 

\multirow{10}{*}{JPetStore} 
& Bunch~\cite{mitchell2008} & 0.10~\cite{saied2024migration} & 2.70~\cite{saied2024migration} & 0.00~\cite{saied2024migration} & 0.38~\cite{saied2024migration} & 4~\cite{saied2024migration} & \Wic 0.05 \\ 
& MEM~\cite{mazlami2017microservice} & 0.17~\cite{saied2024migration} & 3.00~\cite{saied2024migration} & 1.00~\cite{saied2024migration} & 0.43~\cite{saied2024migration} & 9~\cite{saied2024migration} & \Wic 0.04 \\ 
& FoSCI~\cite{fosci2020} & 0.08~\cite{saied2024migration} & 2.80~\cite{saied2024migration} & 0.45~\cite{saied2024migration} & 0.55~\cite{saied2024migration} & 10~\cite{saied2024migration} & \Wic -0.40 \\ 
& CoGCN~\cite{desai2021gnn} & 0.09~\cite{saied2024migration} & 2.00~\cite{saied2024migration} & 0.60~\cite{saied2024migration} & 0.45~\cite{saied2024migration} & 10~\cite{saied2024migration} & \Wic -0.21 \\ 
& Mono2Micro~\cite{kalia2021mono2micro} & 0.05~\cite{kalia2021mono2micro} & 1.86~\cite{kalia2021mono2micro} & 0.33~\cite{kalia2021mono2micro} & \textbf{0.26}~\cite{kalia2021mono2micro} & 10~\cite{saied2024migration} & \Wic -0.24\\ 
& HDBScan~\cite{sellami2022hierarchical} & \textbf{0.22}~\cite{sellami2022hierarchical} & \textbf{0.30}~\cite{sellami2022hierarchical} & \textbf{0.02}~\cite{sellami2022hierarchical} & 0.90~\cite{sellami2022hierarchical} & \ex{-} & \Wic \textbf{0.97} \\ 
& a-BMSC~\cite{saied2024migration} & 0.18~\cite{saied2024migration} & 1.40~\cite{saied2024migration} & 0.65~\cite{saied2024migration} & 0.92~\cite{saied2024migration} & 20~\cite{saied2024migration} & \Wic 0.23 \\ 
& CHGNN~\cite{mathai2021chgnn} & - & - & - & - & - & - \\ 
& MonoEmbed~\cite{sellami2025contrastive} & \ex{0.08} & \ex{2.56} & \ex{0.50} & \ex{0.67} & \ex{9} & \Wic -0.45 \\  \hline 

\multirow{10}{*}{AcmeAir} 
& Bunch~\cite{mitchell2008} & 0.25~\cite{saied2024migration} & 1.50~\cite{saied2024migration} & 0.45~\cite{saied2024migration} & 0.30~\cite{saied2024migration} & 3~\cite{saied2024migration} & \Wic 0.24 \\ 
& MEM~\cite{mazlami2017microservice} & 0.30~\cite{saied2024migration} & 0.70~\cite{saied2024migration} & 0.05~\cite{saied2024migration} & 0.38~\cite{saied2024migration} & 9~\cite{saied2024migration} & \Wic \textbf{0.69} \\ 
& FoSCI~\cite{fosci2020} & 0.33~\cite{saied2024migration} & 1.60~\cite{saied2024migration} & 0.65~\cite{saied2024migration} & 0.70~\cite{saied2024migration} & 9~\cite{saied2024migration} & \Wic 0.19 \\ 
& CoGCN~\cite{desai2021gnn} & \textbf{0.42}~\cite{saied2024migration} & 0.90~\cite{saied2024migration} & 0.60~\cite{saied2024migration} & 1.00~\cite{saied2024migration} & 13~\cite{saied2024migration} & \Wic 0.46 \\ 
& Mono2Micro~\cite{kalia2021mono2micro} & 0.07~\cite{kalia2021mono2micro} & 3.38~\cite{kalia2021mono2micro} & 0.53~\cite{kalia2021mono2micro} & 0.43~\cite{kalia2021mono2micro} & 8~\cite{saied2024migration} & \Wic -0.67 \\ 
& HDBScan~\cite{sellami2022hierarchical} & 0.30~\cite{sellami2022hierarchical} & \textbf{0.30}~\cite{sellami2022hierarchical} & \textbf{0.02}~\cite{sellami2022hierarchical} & 0.90~\cite{sellami2022hierarchical} & \ex{-} & \Wic 0.53 \\ 
& a-BMSC~\cite{saied2024migration} & 0.15~\cite{saied2024migration} & 1.30~\cite{saied2024migration} & 0.80~\cite{saied2024migration} & 0.77~\cite{saied2024migration} & 30~\cite{saied2024migration} & \Wic -0.45 \\ 
& CHGNN~\cite{mathai2021chgnn} & \ex{0.11} & \ex{2.50} & \ex{0.37} & \ex{\textbf{0.00}} & \ex{4} & \Wic -0.15 \\ 
& MonoEmbed~\cite{sellami2025contrastive} & \ex{0.01} & \ex{3.06} & \ex{0.47} & \ex{0.56} & \ex{16} & \Wic -0.85 \\ \hline 
\end{tabularx}} 
\end{table}

Our evaluation of service identification tools across multiple benchmark systems (Table~\ref{tab:combined_results}) reveals several consistent trends 
when considering both the individual metrics and the aggregated \textit{Score} column.
We compare how effectively each tool decomposes a monolithic implementation into microservices, using the score (higher is better) as a summary of structural modularity, interface complexity, communication overhead, and size balance.
Across all datasets, \textbf{HDBScan}~\cite{sellami2022hierarchical} obtains high scores; often ranking first for each benchmark system. This hierarchical clustering approach typically generates a moderate number of services with low interface counts and low inter-partition communication, while keeping NED at acceptable levels. This indicates that, despite their simplicity, hierarchical density–based methods can produce decompositions that strike a good balance between cohesion, coupling, and service-size distribution.
\textbf{a-BMSC}~\cite{saied2024migration} consistently generates the largest number of microservices (e.g., 23 for \texttt{DayTrader} and 30 for \texttt{AcmeAir}), reflecting its bias toward fine-grained partitioning. Its scores are generally moderate: the larger service counts and higher communication (ICP) seem to offset otherwise reasonable structural properties, suggesting that very fine-grained decompositions incur a penalty in our combined metric due to increased coordination overhead.
\textbf{Mono2Micro}~\cite{kalia2021mono2micro} tends to produce comparatively fewer services with competitive scores on \texttt{DayTrader} and \texttt{JPetStore}, reflecting structurally sound but coarser-grained decompositions. However, its performance degrades on some systems, such as  \texttt{Plants} and \texttt{AcmeAir}, where higher IFN and less balanced NED reduce the overall score. This pattern suggests that Mono2Micro favours stable, coarse partitions that can be beneficial for maintainability, but may be suboptimal when finer granularity or strict balance is desired.
\textbf{MEM}~\cite{mazlami2017microservice} and \textbf{CoGCN}~\cite{desai2021gnn} generally achieve mid-range scores. MEM performs particularly well on \texttt{Plants} and \texttt{AcmeAir}, where it combines relatively high SM with moderate IFN and ICP, indicating strong structural clustering when static dependencies are informative. CoGCN shows similar behaviour but is often slightly more penalised by communication and interface complexity, which lowers its aggregate score compared to the best-performing hierarchical approaches.

\textbf{CHGNN}~\cite{mathai2021chgnn} yields mixed and mostly negative scores on the systems where results are available. In our experiments, it produces a relatively small number of services (e.g., 6 for \texttt{DayTrader} and 5 for \texttt{Plants}), which results in higher inter-partition communication and increased interface load per service under our evaluation metrics.
 While such coarser-grained decompositions may be appropriate for certain migration scenarios, they lead to less favorable SM, ICP, and IFN values under our fine-grained comparison setting.

\textbf{MonoEmbed}~\cite{sellami2025contrastive}, which represents monoliths using contrastive-learned embeddings of software components, shows high variability across benchmarks. It attains competitive or positive scores on some systems (notably \texttt{Plants}), where it combines balanced NED with acceptable IFN and ICP, but performs poorly on \texttt{AcmeAir}, where low SM and higher communication drastically reduce its score. Overall, these results indicate that embedding-based methods are promising but highly sensitive to dataset characteristics and hyperparameters.

Overall, our comparison highlights that HDBScan, with its hierarchical clustering strategy, delivers the most consistently strong results across benchmarks. a-BMSC emerges as the second-best performing approach, offering fine-grained decompositions with acceptable communication overhead, while Mono2Micro ranks third, balancing granularity and structural cohesion but incurring higher inter-service interaction in certain cases. MEM exhibits comparable behavior to Mono2Micro, occupying a similar performance tier with distinct trade-offs between service size and coupling. CHGNN and MonoEmbed currently exhibit more dataset-dependent behaviour, implying the need for enhancement.

\section{Limitations}

 The evaluation combines metric values reported in prior work with metrics recomputed under our pipeline, which may affect comparability due to differences in implementations, configurations and  dataset versions, particularly for ICP and IFN, which may be derived from dynamic traces or static approximations. The composite score, while useful for summarization, relies on manually selected weights and contains formulation ambiguities in Equation~\ref{eq:overall-score} that may hinder exact reproducibility; therefore, the resulting rankings should be interpreted as indicative rather than definitive.
The experimental scope also limits generalizability. The evaluation covers only four benchmark systems, was executed on heterogeneous machines, and does not include statistical testing, variance reporting, or sensitivity analysis. Due to limited availability of artifacts and configuration details, full hyperparameter alignment across baselines was not possible, and some tools report incomplete or extreme metric values that could not be independently verified. Moreover, the study does not leverage ground-truth or semi-ground-truth datasets (e.g., TrainTicket) for precision evaluation, nor does it consider cost or runtime performance, which remain directions for future work.

\section{Conclusion}

Our study presented a unified and reproducible evaluation of leading microservice decomposition frameworks, addressing benchmark and metric inconsistencies in prior work. By combining published results with our experiments under a consistent metric pipeline, we enabled a fair comparison of static, dynamic, and hybrid approaches across common benchmark systems.  
Our results show that hierarchical clustering-based methods, particularly HDBScan, consistently produced the most balanced and reliable decompositions, achieving strong structural modularity while maintaining low inter-service communication and interface complexity. a-BMSC and Mono2Micro form a second tier, exhibiting clear trade-offs between fine-grained service autonomy and coordination overhead. 
Classical static-analysis approaches, such as MEM remain competitive on well-structured systems, while representation-learning techniques, including embedding-based and graph neural network methods, display higher sensitivity to dataset characteristics and configuration choices. 

\vspace{-1em}
\setlength{\bibsep}{0pt}
\bibliographystyle{splncs04nat}
\bibliography{references}

\end{document}